# Silicon photonics-integrated time-domain balanced homodyne detector in continuous-variable quantum key distribution

*Yanxiang Jia,*[1,2] *Xuyang Wang,*[1,2,*] *Xiao Hu,*[3,4] *Xin Hua,*[3,4] *Yu Zhang,*[1,2] *Xubo Guo,*[1,2] *Shengxiang Zhang,*[4] *Xi Xiao,*[3,4,*] *Shaohua Yu,*[3,4] *Jun Zou,*[5] *and Yongmin Li*[1,2,*]

[1] State Key Laboratory of Quantum Optics and Quantum Optics Devices, Institute of Opto-Electronics, Shanxi University, Taiyuan 030006, People's Republic of China

[2] Collaborative Innovation Center of Extreme Optics, Shanxi University, Taiyuan 030006, People's Republic of China

[3] State Key Laboratory of Optical Communication Technologies and Networks, China Information and Communication Technologies Group Corporation (CICT), Wuhan 430074, People's Republic of China

[4] National Information Optoelectronics Innovation Center, Wuhan 430074, People's Republic of China

[5] College of Science, Zhejiang University of Technology, Hangzhou 310023, People's Republic of China

## ABSTRACT

We designed and experimentally demonstrated a silicon photonics-integrated time-domain balanced homodyne detector (TBHD), whose optical part has dimensions of 1.5 mm × 0.4 mm. To automatically and accurately balance the detector, new variable optical attenuators were used, and a common mode rejection ratio of 86.9 dB could be achieved. In the quantum tomography experiment, the density matrix and Wigner function of a coherent state were reconstructed with 99.97 % fidelity. The feasibility of this TBHD in a continuous-variable quantum key distribution (CVQKD) system was also demonstrated. This facilitates the integration of the optical circuits of the CVQKD system based on the GG02 protocol on the silicon photonics chip using TBHD.

## INTRODUCTION

Time-domain balanced homodyne detectors (TBHDs) are imperative in quantum information fields, such as quantum tomography[1–5] and continuous-variable quantum key distribution[6–16] (CVQKD). These detectors can be used to measure the quadrature of pulsed quantum signals. Furthermore, TBHDs are compatible with photon counting technology and can be used to detect signal beams with an intensity of approximately several photons. Conventionally, TBHDs are fabricated using free-space optical devices or fiber optical devices.[17,18] With considerable developments in quantum information and large-scale optical integration technology, realizing the conventional functions of time-domain balanced homodyne on an integrated platform is imperative.[19–21]

Recently, frequency-domain balanced homodyne detectors (FBHDs) integrated on silicon photonics chips have been reported.[22,23] These studies focused on the bandwidth of the detectors. However, TBHDs, which use charge and shaping amplifiers, are different from FBHDs, which use a transimpedance amplifier.[24,25] For the optical chips in these studies, vertical coupling was used to couple the light into the chips using a one-dimensional (1D) grating coupler whose insertion loss is > 4 dB. In quantum signal detection, any loss can decrease the signal noise ratio, and edge couplers whose coupling loss can be reduced to approximately 1–2 dB are recommended. To balance 50/50 couplers, the researchers used two 50/50 couplers and a thermal phase modulator between them, which is equivalent to a 2 × 2 Mach-Zehnder interferometer (MZI). [22] The work point of the MZI is on the maximum gradient point and is sensitive to the changes in the drifted bias voltage $V_{bias}$ .Afterward, it becomes unstable. The structure includes one thermal MZI modulator in each output path. [23] The work point of each MZI is on the vertex point, which complicates ensuring the drift direction. The details are shown in Appendix A. Moreover, the insertion loss of a typical MZI is approximately 0.6 dB (each 50/50 multimode interference (MMI) coupler exhibits approximately 0.3 dB loss), which will decrease the detection efficiency of TBHD. In general, the ratio of a real 50/50 coupler is very close to 50/50, and an attenuation of < 3 % is enough to balance it. For the above two kinds of MZI structures, bias voltages $V_{bias}$ should be applied to achieve a well-balanced performance.

The technical challenge to realize TBHD is achieving a high common mode rejection ratio (CMRR) of approximately 70–80 dB stably.[17] The balance of the TBHD requires that the optical intensities and paths of both outputs of the 50/50 coupler are equal, and the response of the two photodiodes (PDs) to the pulsed beam are matched.[17,18] The charge and shaping amplifiers should be designed to ensure that the shot noise to electronic noise ratio (SENR) is high and the output pulse is Gaussian.[14,15] In our study, to realize a well-performing TBHD on a chip, several methods were employed and demonstrated. Firstly, variable optical attenuators (VOAs) based on the PIN phase modulator were used to tune the balance of a 50/50 MMI coupler. The principle of attenuation is free carrier absorption. The insertion loss of the VOA was approximately 0.1 dB, which was less than that of the MZI structures. The size of the PIN phase modulator was also smaller than that of the MZI structures. Secondly, tunable bias voltages of two Ge PDs were used to ensure identical responses. Low-noise charge and shaping amplifiers were used to ensure that a high SENR of 19.42 dB could be achieved. Finally, a CMRR of 86.9 dB is achieved. In the stability test, the measurement time without calibration was 60 s when the SENR was 10 dB. To further demonstrate the performance of the TBHD, a typical quantum tomography experiment was designed and realized. The density matrix and Wigner function of the coherent state $|\alpha\rangle = 1.04$ was reconstructed with 99.97 % fidelity using the maximum likelihood method.

In the quantum communication field, a number of works on integrating the discrete variable quantum key distribution system on silicon photonics-integrated circuits have been conducted.[26-35] Only one work on the silicon photonics integrated CVQKD system was reported.[36] In this study, the continuous beam and FBHD with sideband methods were used. The insertion loss of the receiver chip (using a 2D grating coupler) was high (approximately 6.5 dB). The typical GG02 protocol based on a pulsed beam and TBHD has still not been verified on a silicon photonics chip. Such verification of the GG02 protocol involves several challenges, including generating high-extinction ratio pulses and realizing time multiplexing with delay lines as well as the high CMRR integrated TBHD on a chip. In this paper, not only a silicon photonics-integrated TBHD was realized, but also a 50 km CVQKD experiment based on the TBHD was conducted to demonstrate its feasibility for integration in the system. Our result is expected to pave the way for integrating the whole CVQKD system (TBHD based GG02 protocol) on a chip.

In Section 2, we present the structure and characteristics of the silicon photonics-integrated TBHD. The VOAs are characterized in detail. In Section 3, a quantum tomography experiment based on the silicon photonics-integrated TBHD is demonstrated. In Section 4, CVQKD experiments based on the GG02 protocol are demonstrated with the silicon photonics-integrated TBHD. Finally, Section 5 presents the conclusions.

## STRUCTURE AND CHARACTERISTICS OF SILICON PHOTONICS-INTEGRATED TBHD

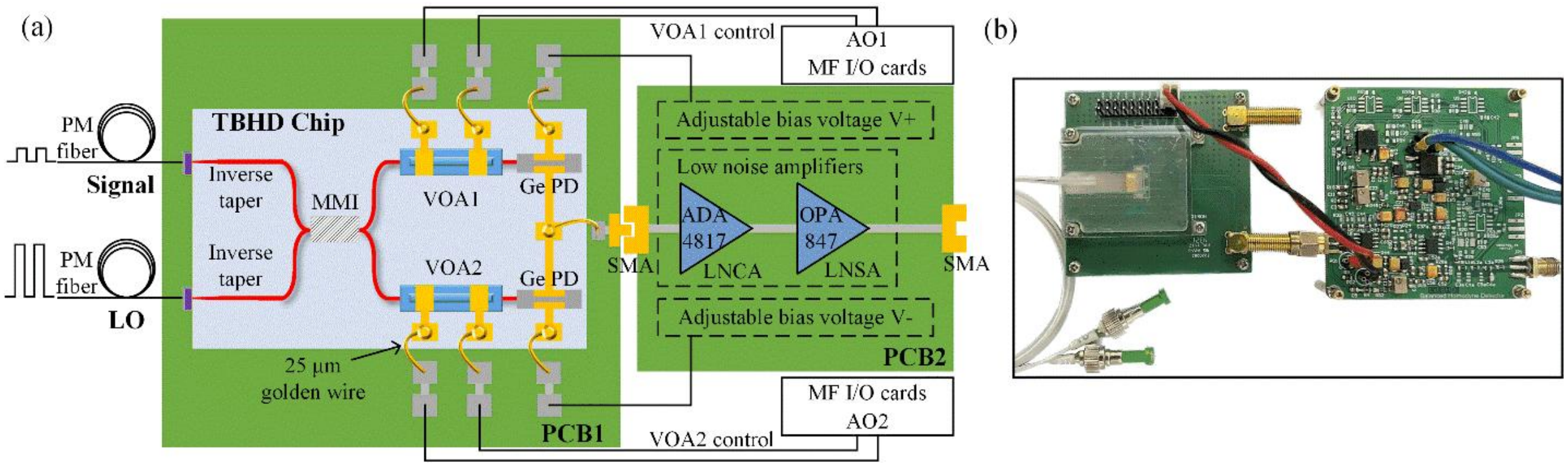

Figure 1. Structure and photograph of silicon photonics-integrated TBHD. (a) Structure of TBHD. (b) Photograph of the TBHD. LNCA: Low-noise charge amplifier; LNSA: Low-noise shaping amplifier; M I/O cards: Multifunctional I/O cards.

The schematic of silicon photonics-integrated TBHD is presented in Figure 1a. The chip was fabricated using industry-standard active flow silicon-on-insulator (SOI) technology. Moreover, it was based on an SOI substrate with a 3-μm Box and 220-nm top silicon. The overall size of the optical part on a chip was 1.5 mm × 0.4 mm, which was much smaller than the fiber-based TBHD. The signal and local oscillator (LO) beams were coupled into the chip through two inverse tapers on the edge of the chip respectively. The guided polarization maintaining (PM) fibers with high numerical aperture values were packaged on the edge of the chip. The size of the inverse taper tip was approximately 140 nm × 220 nm, and the mode field diameter of the inverse taper was around 4.5 μm, which was matched with the numerical aperture of the PM fiber. Subsequently, an edge coupler loss of 2 dB was achieved. In the chip, the signal and LO beams transmitted in TE mode and interfered with each other in a 50/50 MMI coupler. To balance the two output paths of the 50/50 MMI coupler, two VOAs based on PIN phase modulators were used, which introduces excess losses owing to the free carrier absorption effect when forward voltages were applied. At the end of the two output paths, two Ge PDs were utilized to detect the interfered beams. The responsivity was 0.9 A/W, and the dark current was 64 nA at 3.7 V. To balance the optical paths of the two output paths, the lengths of the two output waveguides of the 50/50 MMI coupler were designed to be the same. The bias voltages of two Ge PDs were tunable to ensure that the responses of the two PDs were identical. Based on the above methods, the two output pulsed photoelectrons are well-balanced and subtracted sufficiently.

The low-noise charge and shaping amplifiers following the chip are critical to ensure that the TBHD operates at a low electronic noise level. The Op Amps ADA4817 was used to integrate subtracted photoelectrons and served as the low-noise charge amplifier. For the low-noise shaping amplifier, OPA847 was selected for its high bandwidth. Further details about the electronic circuits can be found in (17), (18). The whole setup was enclosed in a metal box to resist electromagnetic interference (Figure 1b).

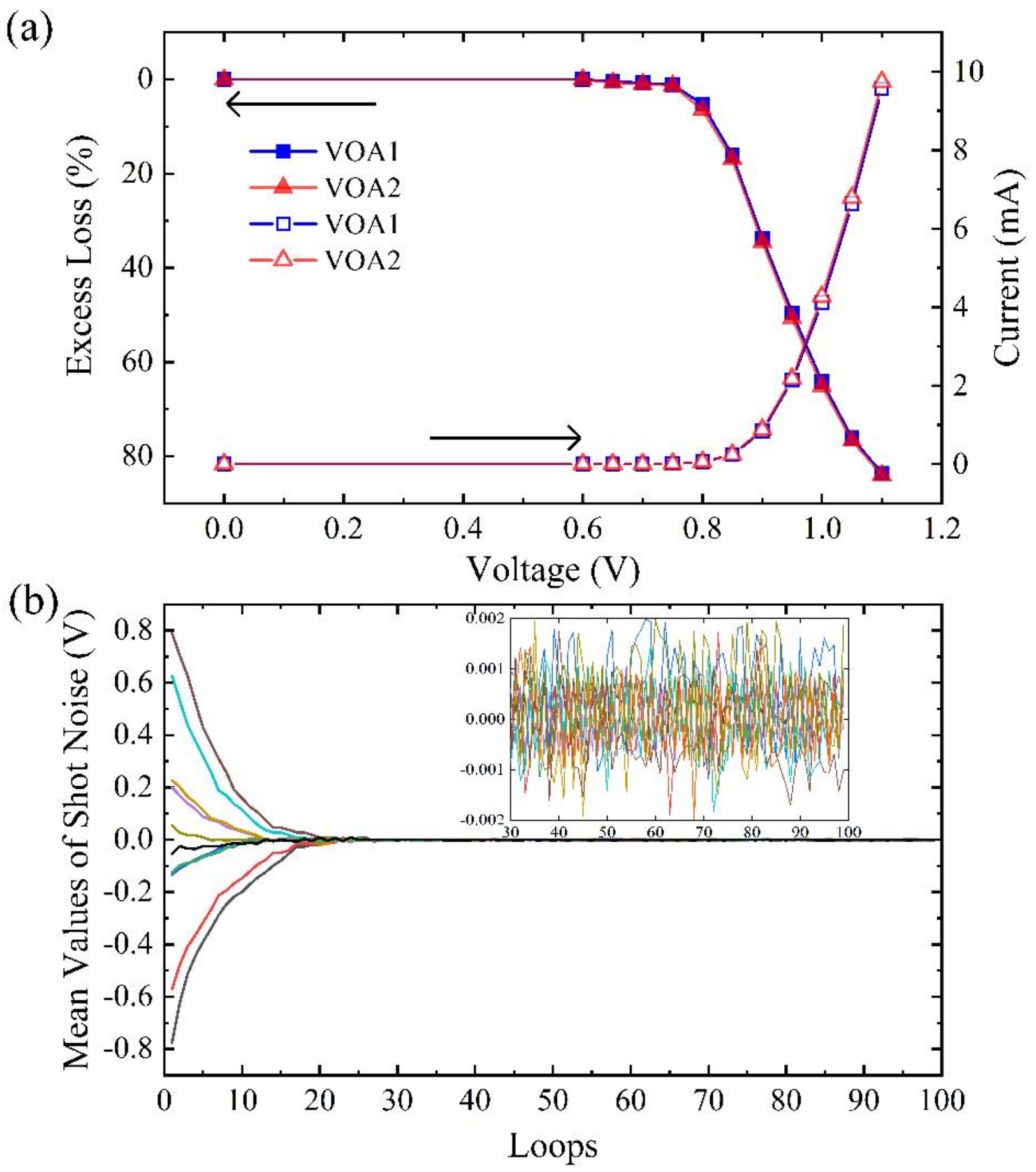

Figure 2. (a) Excess loss against the applied forward voltage of the PIN phase modulator and the corresponding voltage–current attribution line. (b) Evolution process of the detector from unbalanced to balanced states.

Figure 2 depicts the excess losses against the applied forward voltages on the PIN phase modulators. The solid blue square and red triangle lines represent excess losses of the two VOAs, and the hollow square and triangle lines are the corresponding voltage–current attribution lines. The maximum attenuated amplitude was > 80%, and the corresponding current was approximately 9.5 mA, which is below the damage threshold of approximately 10 mA. The attenuated amplitude was enough to compensate for the imbalance of the two output paths of the 50/50 MMI coupler, which has a typical imbalance lever of < 3%.[37]

Table 1. The excess losses and currents against the applied voltages

| Forward Voltage (V) | VOA1 | | VOA2 | |
|---|---|---|---|---|
| | Loss (%) | Current (mA) | Loss (%) | Current (mA) |
| 0 | 0.0 | 0 | 0.0 | 0 |
| 0.6 | 0.0 | 0 | 0.0 | 0 |
| 0.65 | 0.4 | 0 | 0.7 | 0 |
| 0.7 | 0.7 | 0 | 1.0 | 0 |
| 0.75 | 1.1 | 0.01 | 1.3 | 0.01 |
| 0.8 | 5.3 | 0.06 | 6.5 | 0.06 |
| 0.85 | 16.0 | 0.24 | 16.9 | 0.24 |
| 0.9 | 33.7 | 0.84 | 34.5 | 0.88 |
| 0.95 | 49.5 | 2.14 | 50.7 | 2.18 |
| 1 | 64.1 | 4.12 | 65.1 | 4.27 |
| 1.05 | 76.1 | 6.64 | 76.7 | 6.79 |
| 1.1 | 83.6 | 9.56 | 84.0 | 9.74 |

Table 1 lists the values of all points in Figure 2. When the applied voltage was between 0.65 and 0.7 V, an average voltage precision of approximately 1.67 mV was required to reach the tuning sensitivity of $1 \times 10^{-4}$. When the applied voltage was between 0.75 and 0.8 V, an average voltage precision of approximately 0.12 mV was required to reach the tuning sensitivity of $1 \times 10^{-4}$. Thus, voltage precision increases with the applied voltage. Based on the excess loss–voltage curve, an auto-balancing technique can be designed. The technique is similar to the auto-balancing technology based on the loss-radius behavior of the bent fiber.38 The auto-balancing processes and results are shown in Figure 2b. Approximately 20 loops were required from any initial unbalanced state to the balanced state. When the intensities of two output paths of the 50/50 MMI coupler were balanced, the mean values of the shot noise could be controlled in the ± 2 mV range.

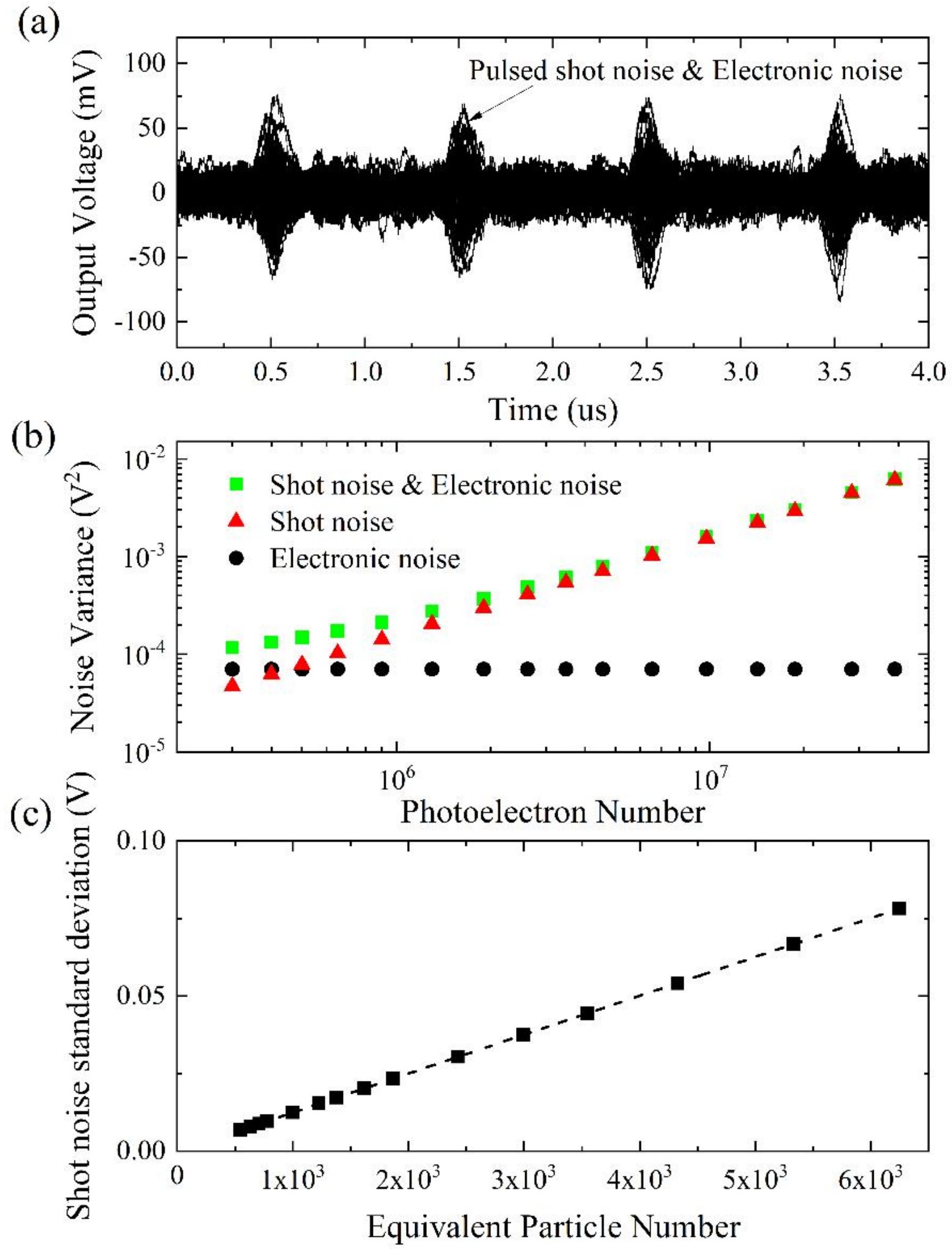

Figure 3. Performance of TBHD. (a) Time traces of output electronic pulse. (b) Variances of the electronic and shot noises and their sum versus the photoelectron number. (c) Linear gain of TBHD.

With the above auto-balancing technique, the oscilloscope (Tektronix MSO 5204B) traces of the output shot noise are shown in Figure 3a. The pulse repetition rate is 1 MHz, and the full width of the output electric pulse is approximately 200 ns. The measured photoelectron number of the two PDs is $1.42 \times 10^7$, and the SENR is 15 dB. A CMRR of 86.9 dB could be measured. The detailed measurement method of CMRR of TBHD is shown in Appendix B. In comparison to the TBHD with commercial InGaAs photodiodes, selecting the matched photodiodes was unnecessary. We infer that the two Ge PDs were fabricated in the same wafer using the same procedures; therefore, they exhibited a similar performance.

To measure the linearity of the detector, the intensity of the LO beam was tuned and the measurement result was shown in Figure 3b. The green square points represent the measured variance, which is the sum of the shot and electronic noises. After subtracting the electronic noise variances (black circle points, $7 \times 10^{-5}$ $V^2$), the remaining variances shown as red triangle points represent the shot noise variance. A maximum SENR of 19.42 dB is achieved. The relationship between the shot noise variance $\Delta V_{\mathrm{SN}}^2$ and photoelectron number $\eta \cdot |\beta|^2$ can be expressed as follows:

$$\Delta V_{\mathrm{SN}}^2 = G^2 \cdot \eta \cdot |\beta|^2 \quad (1)$$

where $\eta$ is the detection efficiency of the TBHD, and $|\beta|^2$ is the LO beam intensity. From eq 1, the gain of TBHD can be calculated as follows:

$$G = \sqrt{\Delta V_{\mathrm{SN}}^2} \Big/ \sqrt{\eta \cdot |\beta|^2} \quad (2)$$

The gain is also the slope of the fitting dash line in Figure 3c. The horizontal and vertical coordinate values of the black points are the square roots of the coordinate values of the red points in Figure 3b. The term $\sqrt{\eta \cdot |\beta|^2}$ can be defined as the equivalent particle number. The gain of the detector was $1.25 \times 10^{-5}$ V/photon, and the sum of squares of residual value was $2.27 \times 10^{-36}$ $V^2$ in the linear fitting, indicating that TBHD has good linearity.

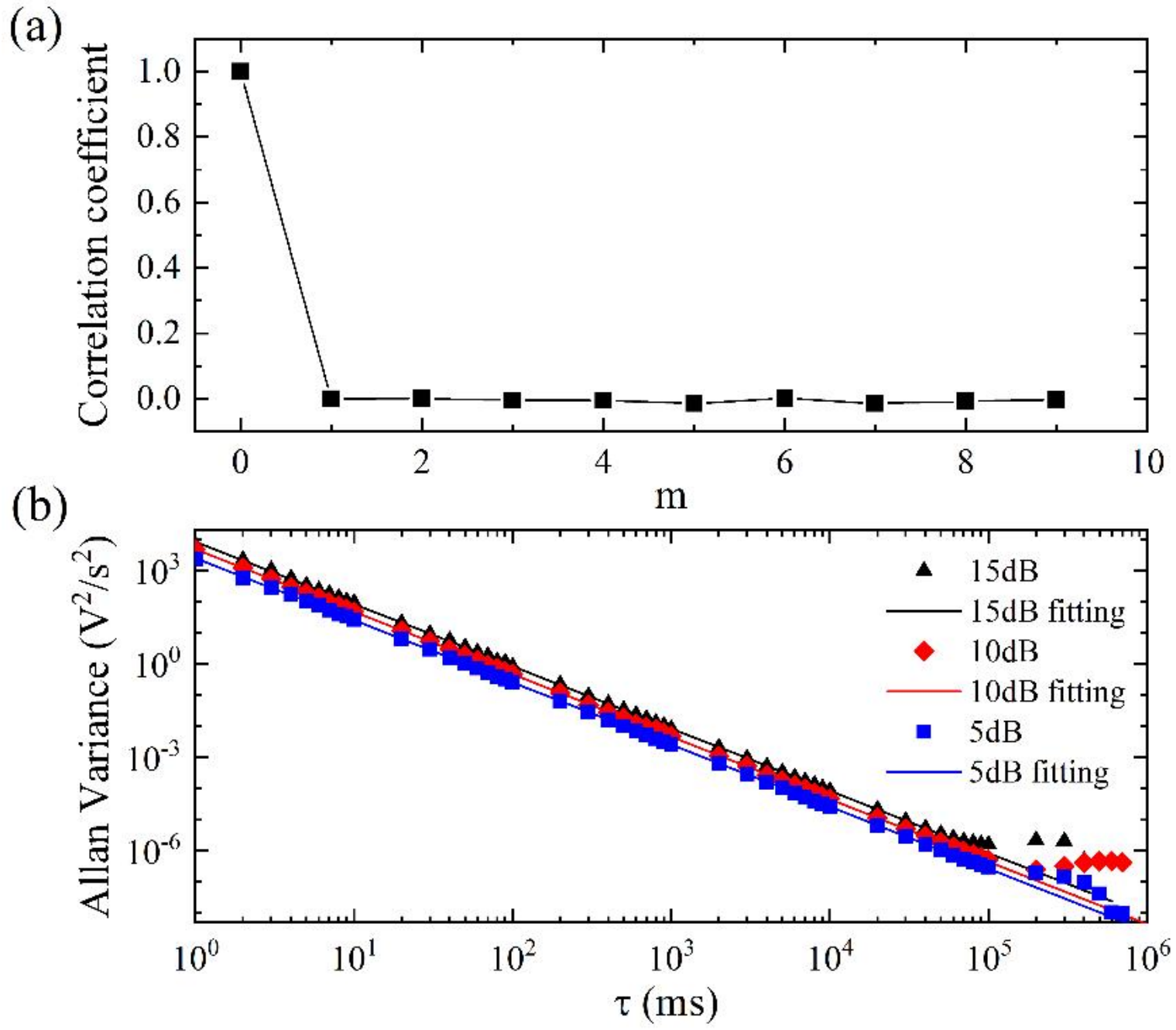

Figure 4. Time-resolving power and stability of TBHD. (a) Correlation coefficient (CC) between n-th and (n + m)-th pulses. (b) Allan variance of the TBHD peak value in a period of $10^6$ ms.

The correlation coefficient (CC) between adjacent pulses was calculated using eq 3 to check the time-resolving power of TBHD. In the experiment, approximately $5 \times 10^6$ output peak points $P(i)$, $(i = 1, 2, \cdots, 5\times10^6 + 9)$ were used.

$$\mathrm{CC}\left(m\right)=\frac{\left\langle XY\right\rangle-\left\langle X\right\rangle\left\langle Y\right\rangle}{\sqrt{\left\langle X^2\right\rangle-\left\langle X\right\rangle^2}\sqrt{\left\langle Y^2\right\rangle-\left\langle Y\right\rangle^2}} \tag{3}$$

where $X(i) = P(i)$ and $Y(i) = P(i+m)$, $(i = 0, 1, \cdots, 5\times10^6;\ m = 0, 1, 2, \cdots, 9)$ . Figure 4a presents the CC when the pulse repetition rate is 1 MHz. A perfect correlation occurred when $m = 0$ ,

whereas the correlation vanishes for $m \neq 0$ . By evaluating the CC, we confirmed that the detector could distinguish individual optical pulses at a repetition rate of 1 MHz.

The stability of TBHD was essential for determining the duration for which the detector could work accurately without calibrating the low-frequency drift of the detector's balance. The total photoelectron number of two Ge PDs is 4.55 × $10^6$ when the SENR was 10 dB, and the corresponding standard deviation of the shot noise was 26.6 mV. Considering the gain of the detector was 1.25 × $10^{-5}$ V/photoelectron, a one-in-a-thousand drift of detector balance could cause a drift of 56.8 mV, which was about 2.1 shot noise units. Thus, a slight imbalance of the TBHD could result in a significant deviation of the output signal. The detector's stability at various time scales was analyzed using Allan variance, which is defined as follows:

$$\sigma^2(n\tau_0, N) = \frac{1}{2n^2\tau_0^2(N-2n)} \cdot \sum_{i=0}^{N-2n-1} (x_{i+2n} - 2x_{i+n} + x_i)^2 \tag{4}$$

where $x_i$ is the measured quadrature value, $\tau_0$ ( $\tau_0 = 1\text{ms}$ ) is the time interval between the adjacent sampling points, $n$ is the number of sampling points over time, and $N$ is the total number of samples. All the quadrature values were acquired at 1kHz at a time scale of $10^6$ ms. Afterward, different $n$ were used to calculate the Allan variance at different time scales, as shown in Figure 4b. The solid lines were linearly fitted using the data points for time < 1 s. The stability at different SENRs was measured. The time windows for accurate measurement without calibration were 50, 60, and 70 s when the SENRs were 15, 10, and 5 dB, respectively. We found that the detector shows better stability if the SENRs are lower.

From the aforementioned analysis, we find that the silicon photonics-integrated TBHD performs well. Its performance is comparable to the TBHD with commercial InGaAs photodiodes, except that the detection efficiency is lower (mainly owing to the 2 dB edge coupling loss) and the dark current is larger. In the future, the dark current Ge PD and edge coupling loss are expected to be improved. Fortunately, there is no need to select the matched photodiodes for TBHD on a chip. In the following section, quantum tomography and CVQKD experiments using the silicon photonics-integrated TBHD will be described.

## CHIP TBHD IN TOMOGRAPHY

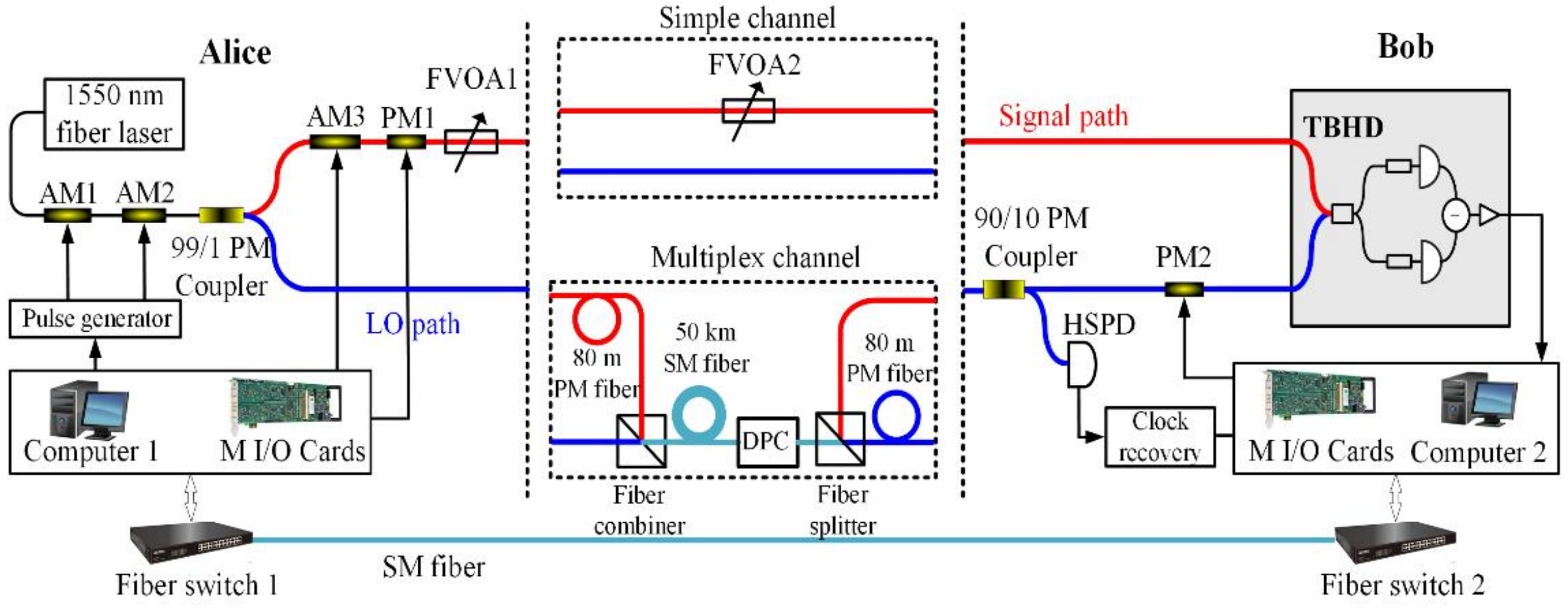

Figure 5. Scheme of the setup to perform quantum tomography and CVQKD. HSPD: high-speed photodetector; M I/O cards: Multifunctional I/O cards.

As an application of our TBHD detector, we performed quantum tomography of a coherent state by reconstructing its density matrix and Wigner function. Figure 5 presents the scheme for performing quantum tomography using a simple channel. This was also a simple CVQKD setup used to evaluate the excess noise of the system.[39] At the sender, Alice's side, a continuous 1550-nm laser beam (Gooch & Housego, EM650) was stably modulated into 80-dB high-extinction ratio pulsed beams by two series amplitude modulators (AM1 and AM2, iXblue, MXER-LN-10).[40] The pulse width was 100 ns, and the repetition rate was 0.5 MHz. Subsequently, they were split into signal and LO beams by a 99/1 PM fiber coupler. The signal beam could be modulated in the phase space using the amplitude modulator AM3 and phase modulator PM1 (KEYANG PHOTONICS, KY-PM-15-300M). The fiber VOAs FVOA1 and FVOA2 could attenuate the intensity of the signal beam to a suitable intensity range. In the simple channel, the signal or LO beam on the sender side was directly connected to the signal or LO beam on the receiver side. At the receiver, Bob's side, a 90/10 PM fiber coupler and high-speed photodetector were used to recover the clock signal. Moreover, phase modulator PM2 was used to lock the relative phase and shift the measurement base randomly. The signal and LO beams were guided into the silicon photonics-integrated TBHD. A computer and related multifunction I/O cards were presented on each side. Two computers were connected by fiber switches, which were used to transmit classical information.

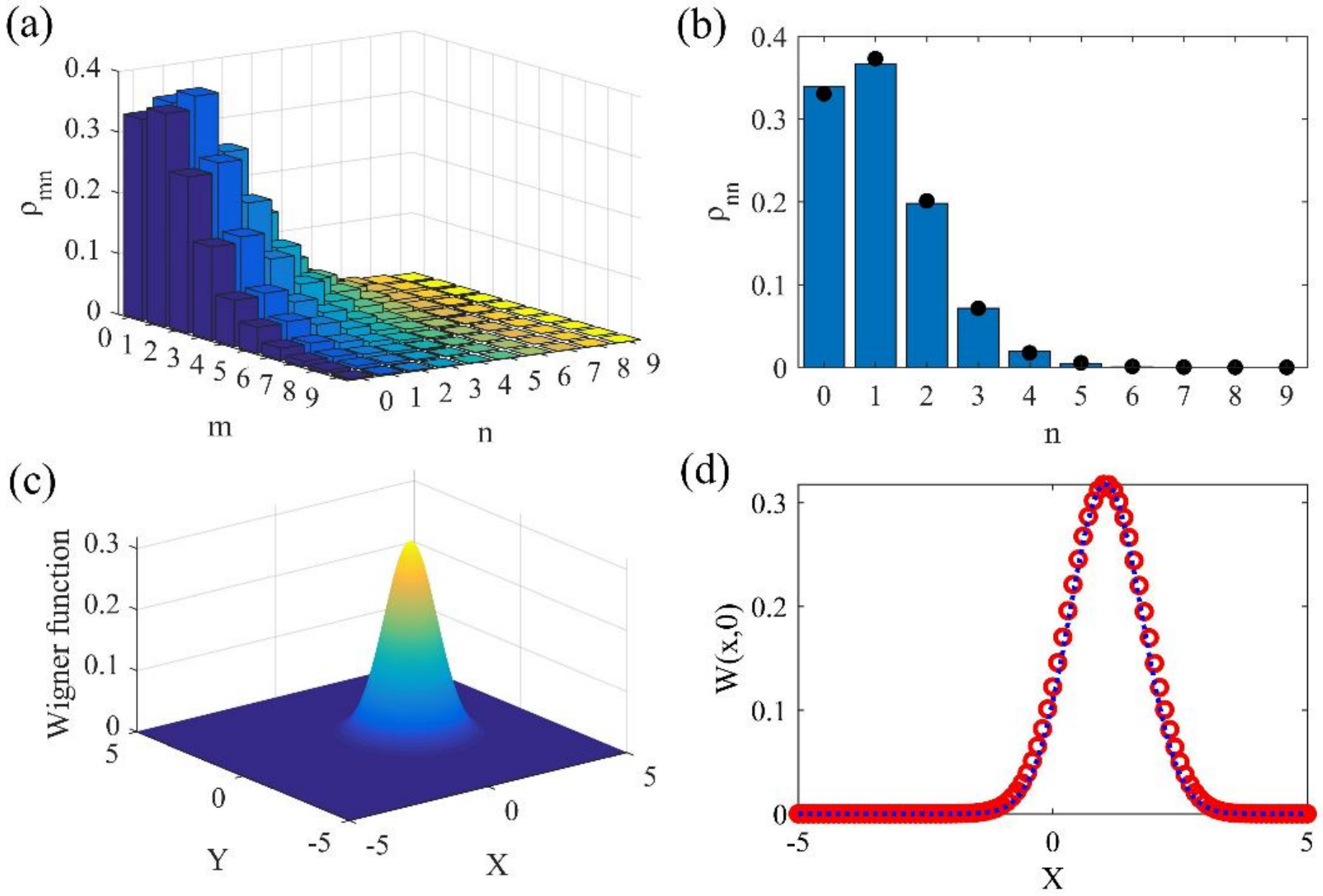

Figure 6. Reconstructed density matrix and Wigner function of the coherent state using the maximum likelihood method. (a) Reconstructed density matrix. (b) Measured photon number

distribution (black dots), the Poisson distribution of coherent state $|\alpha\rangle = 1.04$ (blue columns) is shown for comparison. (c) Reconstructed Wigner function. (d) The reconstructed W (x; 0) (red circles) and W (x; 0) of the coherent state $|\alpha\rangle = 1.04$ (black dash line).

In the experiment, the sender Alice accurately modulated the signal beam intensity using AM3 and FVOA1 (FVOA2 is set to 0 dB attenuation). Subsequently, the phase was scanned by PM1 from 0 to 2π with an interval of 2π/5000. In each phase, one quadrature value x was acquired. Overall, 5000 samples were used to reconstruct the density matrix $\hat{\rho}$ with the iteration operator $\hat{R}$ using the maximum likelihood method.[1] Figure 6a presents the reconstructed density matrix. The iteration expression is shown in eq 5.

$$\hat{\rho}^{(k+1)} = N\left[\hat{R}\left(\rho^{(k)}\right) \cdot \rho^{(k)} \cdot \hat{R}\left(\rho^{(k)}\right)\right] \tag{5}$$

Using the reconstructed density matrix, an average photon number $\langle n\rangle = \sum n \cdot \rho_{nn} = 1.08$ was calculated from the measured photon number distribution. It corresponds to the coherent state $|\alpha\rangle = 1.04$ . The comparison of the reconstructed density matrix with that of an ideal coherent state $|\alpha\rangle = 1.04$ yields a state preparation fidelity $F$ :

$$F = \langle \alpha \,|\, \hat{\rho} \,|\, \alpha \rangle = 99.97\% \tag{6}$$

Figure 6b presents the comparison of measured photon number distribution with a Poisson distribution of coherent state $|\alpha\rangle = 1.04$ . Considering the detection efficiency $\eta = 0.38$ at the receiver's side. The coherent state prepared by Alice is $|\alpha\rangle = 1.69$ .

Based on the calculated density matrix $\hat{\rho}$, the Wigner function can be calculated using the following equation:

$$W(x,y) = tr\left[\hat{\rho}\cdot\exp\left[-2\left(\hat{a}^{\dagger}-\gamma^{*}\right)\left(\hat{a}-\gamma\right)\right]\Big/\pi\right] \quad (7)$$

where $\hat{a}^{\dagger}$ and $\hat{a}$ are the creation and annihilation operators, and $\gamma$ is defined as $\gamma = x + iy$. Figure 6c presents the reconstructed Wigner function. In Figure 6d, the red circle line presents the reconstructed $W(x,0)$, which coincides well with $W(x,0)$ of the coherent state $\alpha = 1.04$. From the aforementioned results, we can infer that the silicon photonics-integrated TBHD could reliably reconstruct the quantum state.

CHIP TBHD IN CVQKD EXPERIMENT

The main aim of the CVQKD experiment was to demonstrate that excess noise could be measured within a reasonable region using silicon photonics-integrated TBHD. The experiment setup is shown in Figure 5.

In the CVQKD experiment, the GG02 protocol was utilized, and all the variances were normalized to the shot noise. At the sender side, the Gaussian modulation variance was calibrated as $V_A = 5.8$, and the electronics noise was $\upsilon_e = 0.1$. Moreover, a reverse reconciliation efficiency of $\beta = 0.98$ was used. The data block containing data samples of $10^8$ was used to evaluate the parameters and extract the secret keys. In the experiment, transmission losses of 5 and 10 dB (set by FVOA2) in the simple channel and 50 km in the multiplexing channel were used. In the multiplexing channel, 80-m PM fibers were designed for time multiplexing, and the fiber combiner and splitter were used for polarization multiplexing. The multiplexing channel was used to prevent the interference of the signal and LO beams and the fast change of the relative phase between both beams. The dynamic polarization controller (DPC, General Photonics, PCD-M02-4X-7-FC/APC) was used to rotate the polarization of the combined signal and LO pulses back to linear polarization, and then they could be split by the fiber splitter. The excess noises and corresponding secret key rates of the CVQKD system were measured and depicted in Figure 7.

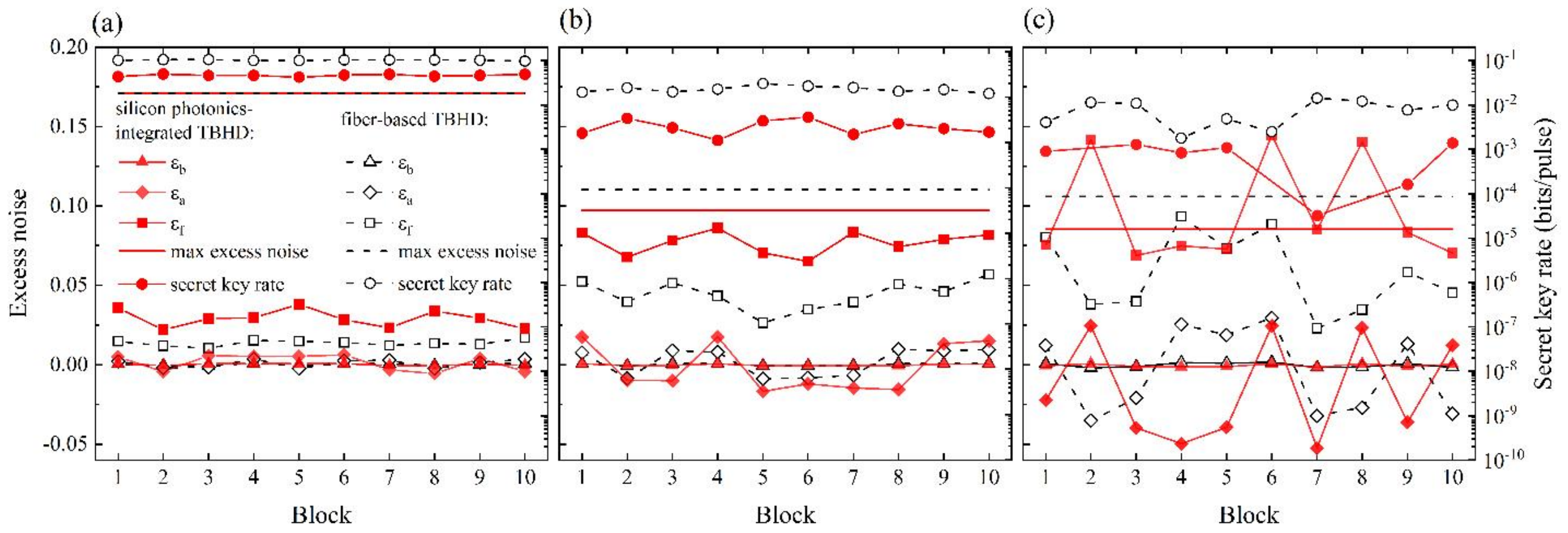

Figure 7. Excess noises and secret key rates in different transmission situations. (a) Transmission loss of 5 dB and (b) 10 dB in a simple channel. (c) Fifty-kilometer single-mode fiber in the multiplexing channel.

In Figure 7, the red solid points represent the experimental results of the CVQKD system based on the silicon photonics-integrated TBHD. For comparison, the black hollow points with the same shape represent the experiments conducted on the fiber-based TBHD with a commercial InGaAs photodiode. The calibrated detection efficiency was 0.72. Figure 7a depicts the case in which the total transmission loss was 5 dB in the simple channel. The solid red triangle represents the excess noise $\varepsilon_b$ at Bob's side using the following equation[41]:

$$\varepsilon_b = \left\langle x_B^2 \right\rangle - 1 - \upsilon_e - T\eta V_A \tag{8}$$

where $\left\langle x_B^2 \right\rangle$ represents the variance of the measured quadrature at the receiver. $T$ represents the transmission loss and $\eta$ is detection efficiency. They nearly overlapped with the hollow black triangle points. There was negative excess noise of $\varepsilon_\mathrm{b}$, mainly due to the statistical effect of finite samples. The solid red diamond points represent the excess noise $\varepsilon_a = \varepsilon_b / \eta T$ normalized at the input port of the transmission channel. The solid red square points represent the excess noise $\varepsilon_f$ considering the finite-size effect:

$$\varepsilon_f = \varepsilon_a + z_{\delta_{PE/2}} \frac{\left(1 + \eta T \varepsilon + \upsilon_e\right)\sqrt{2}}{\eta T \sqrt{m}} \tag{9}$$

where $z_{\delta_{PE/2}} = \sqrt{2} erf^{-1}\left(1 - \delta_{PE}\right)$, and $m$ is the amount of data used to estimate the parameters. Each data block contains $N = 10^8$ data points. The optimum ratio of data used to evaluate the parameters is $r = m / N$, which is 0.1. The final secret key rate of the system is calculated using eq 10. The results are drawn as red solid circle points.

$$K_r^f = \frac{n}{N}\left(\beta I_{AB} - \chi_{BE}^{\delta_{PE}} - \Delta(n)\right) \tag{10}$$

where $n = N - m$ represents the number of signals used to extract the secret keys. $I_{AB}$ is the Shannon mutual information between Alice and Bob. $\beta$ is the reverse reconciliation efficiency. $\chi_{BE}^{\delta_{PE}}$ represents the maximum of the Holevo information compatible with the statistics, except with the probability $\delta_{PE} = 10^{-10}$. Moreover, $\Delta(n)$ is a correction term for the achievable mutual information in the finite case.[42,43]

Figure 7b shows the excess noises and secret key rates when the transmission loss was 10 dB in the simple channel. Figure 7c plots the corresponding excess noise and secret key rates when the transmission channel was 50 km fiber in the multiplexed channel. Considering the DPC and fiber splitter losses, the detection efficiency of the receiver TBHD on the chip changed to 0.28. Correspondingly, the detection efficiency of fiber-based TBHD changed to 0.53. The red solid and black dashed lines in each subfigure represent the maximum excess noise that can be endured by the CVQKD system. The excess noise $\varepsilon_f$ of some blocks exceeded the maximum allowable excess noise level (Figure 7c), and their secret key rates drop to zero. The excess noises $\varepsilon_{\mathrm{b}}$ in the three situations were nearly the same. The excess noises $\varepsilon_a$ were different mainly owing to the difference in the transmission loss $T$ and detection efficiency $\eta$. For example, the excess noise using silicon photonics-integrated TBHD was approximately double that of the fiber-based TBHD in the simple channel, mainly because the detection efficiency of the former was half that of the latter. Thus, the performances of the two kinds of TBHD were similar in evaluating the excess noise except for the detection efficiency.

From the aforementioned results, we can conclude that the silicon photonics-integrated TBHD was suitable for the CVQKD system. Additionally, the excess noise could be controlled to a low level, as in the CVQKD system with fiber-based TBHD. Because the excess noise $\varepsilon_a$ was normalized to the input port of the transmission channel, it increased when the value of $\eta T$ decreased. Furthermore, the coupler loss and quantum efficiency that affect the detection efficiency should be further optimized to improve the performance of the TBHD on a chip.

## CONCLUSION

We designed and demonstrated a silicon photonics-integrated TBHD based on industry-standard SOI technology. The structures and characteristics are presented in detail. The signal and LO oscillator beams could transmit into the chip using packaged edge coupler technology with an insertion loss of approximately 2 dB. Using VOAs based on the PIN phase modulator and Ge photodiodes with tunable bias voltage, a CMRR of 86.9 dB could be achieved. The charge and shaping amplifiers used the low-noise ADA4817 and OPA847 to ensure that a maximum SENR of 19.42 dB could be achieved. The TBHD could output a Gaussian-shaped pulse with a full pulse width of 200 ns. The mean value of the shot noise could be controlled in the range of ± 2 mV automatically. The detector had good linearity, and a gain of 12.5 μV/photoelectron. Using the TBHD, a quantum tomography of a coherent state $\left|\alpha\right\rangle = 1.04$ was conducted, and 99.97 % fidelity was achieved. To demonstrate the performance of the integrated TBHD in the CVQKD system, the CVQKD setup with both simple and multiplexing channels was designed. When the transmission channel was 5 dB loss, 10 dB loss, and 50 km fiber, the excess noise could be controlled to 0.029 ± 0.009, 0.077 ± 0.012, and 0.096 ± 0.049, respectively, and positive secret key rate can be achieved considering the finite-size effect. The stability of TBHD at different SENRs was measured, and the detector showed better stability if the SENR was lower.

Our results indicate that the silicon photonics-integrated TBHD could perform as well as the TBHD based on the commercial InGaAs photodiode. Because the two photodiodes were fabricated in the same wafer using the same procedures, their characters were considerably similar. Therefore, a high CMRR could be achieved, and there was no need to select PDs on chips. For further study, it was expected that the whole GG02 protocol based CVQKD system could be integrated into the chip. In the receiver, the main challenge was integrating the TBHD on the chip, which has been verified in our work. At the sender, the main difficulty was generating high-extinction ratio pulses. For the whole system, time multiplexing was also a challenging technology, as delaying hundreds of nanoseconds (tens of meters optical path) on the chip was difficult. One convenient method was connecting the delayed fibers to the chip through an arrayed edge coupler. The other method was shorting the pulse width to nanoseconds (tens of centimeters optical path) or tens of picoseconds (tens of micrometers optical path). In this way, time multiplexing could be realized on the chip.

Overall, we designed and demonstrated silicon photonics-integrated TBHD, which performed well in quantum tomography and CVQKD experiments. Our results will pave the way for integrating the entire setup of a CVQKD system based on the GG02 protocol using TBHD into a silicon photonics chip.

APPENDIX A: THE MZI STRUCTURES OF BALANCING THE FBHD DETECTORS

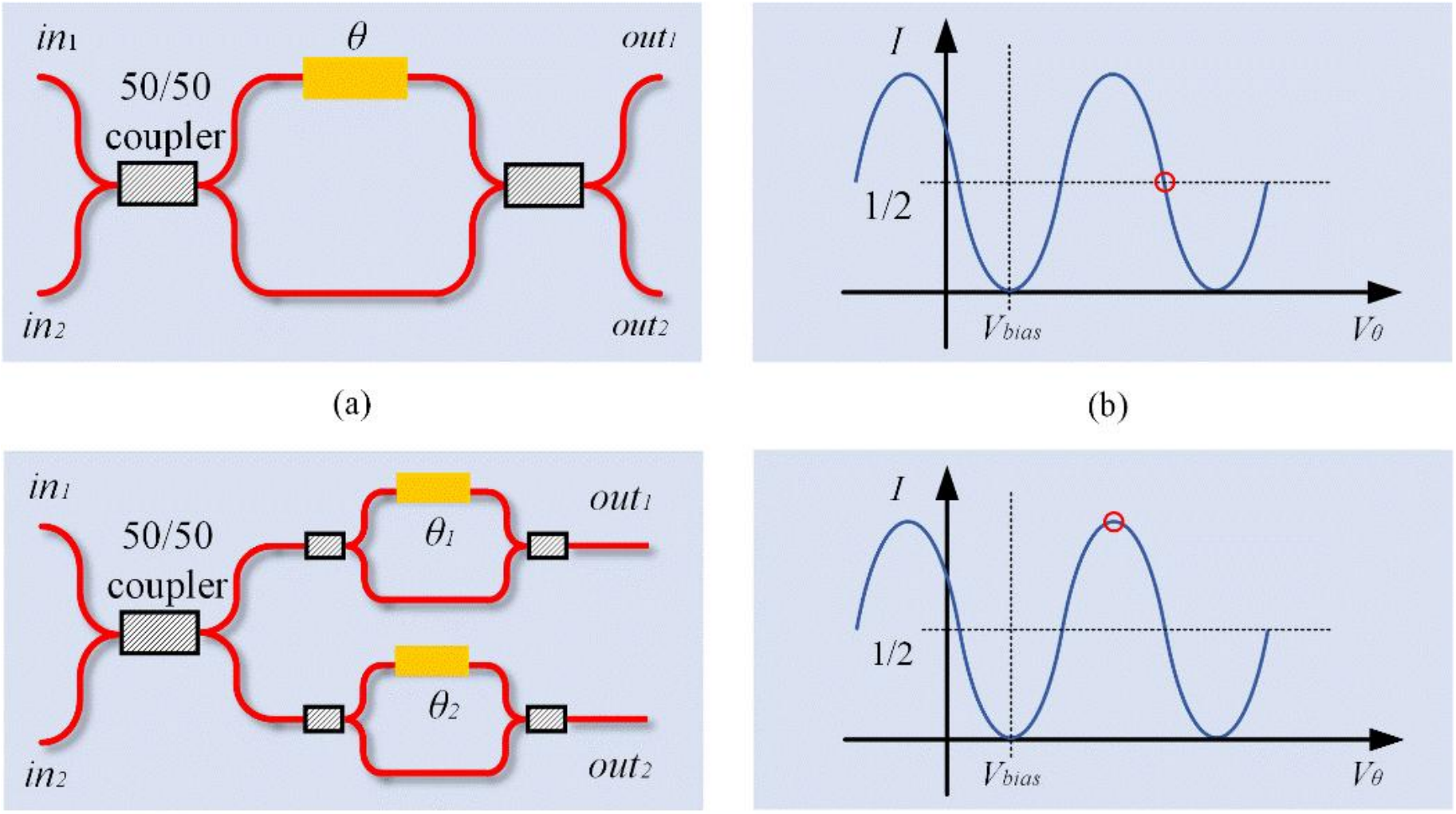

(a) (b)

(c) (d)

Figure 8. Two optical structures of BHD on a chip. (a) One 2 × 2 MZI structure. (b) The output intensity versus applied voltage. (c) Two 1 × 1 MZI structures. (d) The output intensity versus applied voltage.

The optical structure of FBHD which shown in Figure 8a is a typical 2 × 2 MZI.[23] The input ports are $in_1$ and $in_2$, and the output ports are $out_1$ and $out_2$. When a laser beam is injected into one input port of the MZI, the normalized output light intensity is given by:

$$I = \frac{1 \pm \cos\theta}{2} \tag{A1}$$

The relative phase $\theta$ can be controlled by applied voltage $V_\theta$. Considering the bias voltage $V_{bias}$, the intensities $I$ can be transformed into the following:

$$I = \left(1 + \cos\left[\left(V - V_{bias}\right) \cdot \pi / V_\pi\right]\right) / 2 \tag{A2}$$

where $V_{bias}$ corresponds to biased voltage of the relative phase $\theta_{bias}$. When the light intensities of the output port $out_1$ and $out_2$ are balance, the relative phase $\theta$ is $\pi/2 \pm n\pi$. The work point is around the red circle in Figure 8b, which has the maximum gradient. The balance point is very sensitive to the drifted $\theta_{bias}$ or $V_{bias}$.

Figure 8c presents the optical structure with each output path of the 50/50 MMI coupler, a 1 × 1 MZI was used to tune the balance of the two output.[23] The normalized light intensity of each output port can be represented as eq A1. The work point is around the red circle (vertex point) in Figure 8d. When the $V_{bias}$ drifts, it is hard to determine which direction to tune the voltage,

For the above two kinds of MZI structures, bias voltages $V_{bias}$ should be applied and related locking techniques are needed to achieve a precise and stable balanced performance.

APPENDIX B: MEASUREMENT METHOD OF CMRR OF TBHD

The measurement of CMRR of TBHD differs from that of FBHD. The CMRR of the FBHD can be measured using a radio frequency spectrum analyzer directly with sideband method.[44] However, this method is not suitable for TBHD. The CMRR of TBHD can be calculated by the following equation:

$$CMRR = 20 \cdot \log_{10} \left| \frac{G_{\mathrm{DM}}}{G_{\mathrm{CM}}} \right| \tag{B1}$$

where $G_{\mathrm{DM}}$ is the differential mode magnification, and $G_{\mathrm{CM}}$ is the common mode magnification. $G_{\mathrm{DM}}$ is also the gain of the charge and shaping amplifiers, and it can be calculated using the following equation:

$$\Delta V_{\mathrm{SN}}^2 = G_{\mathrm{DM}}^2 \cdot \eta \cdot |\beta|^2 \tag{B2}$$

where $|\beta|^2$ is the LO beam intensity, $\eta|\beta|^2$ is the photoelectron number and $\Delta V_{SN}^2$ is the shot noise variance. When we measure $G_{\mathrm{CM}}$, the two output paths of TBHD should be tuned to balance. However, even if the light intensities of two paths are equivalent, the responses of two photodiodes cannot be exactly the same. Thus, the remaining photoelectrons $\Delta$ will generate a voltage $\overline{V}_{SN}$ which is the non-zero mean value of shot noise. Their relationship can be expressed as:

$$\Delta V_{SN} = G_{DM} \cdot \Delta \tag{B3}$$

The relationship of the mean voltage $\bar{V}_{SN}$ and input photoelectron $\eta \cdot |\beta|^2$ can be expressed as follows:

$$\bar{V}_{SN} = G_{CM} \cdot \eta \cdot |\beta|^2 \tag{B4}$$

In the experiment, the photoelectron number was $4.82 \times 10^6$/pulse, and the SENR was 10.3 dB. The mean value $\bar{V}_{\mathrm{SN}}$ of the output variance was 0.00276 V. The gain $G_{\mathrm{CM}}$ could be calculated as $5.72 \times 10^{-10}$. Finally, the CMRR could be calculated as 86.9 dB.

AUTHOR INFORMATION

**Corresponding Author**

* E-mail: wangxuyang@sxu.edu.cn.

* E-mail: xxiao@wri.com.cn

* E-mail: yongmin@sxu.edu.cn.

**ORCID**

Xuyang Wang: 0000-0003-0055-0929

Xi Xiao:0000-0003-3696-877X

Yongmin Li: 0000-0003-0228-7693

**Funding Sources**

We acknowledge the funding from the Provincial Natural Science Foundation of Shanxi, China (202103021224010), Research Project Supported by Shanxi Scholarship Council of China (2022-016), Key Research and Development Program of Guangdong Province

(2020B0303040002), Aeronautical Science Foundation of China (20200020115001); National Natural Science Foundation of China (62175138), Shanxi 1331KSC, and Open Project of the State Key Laboratory of Quantum Optics and Quantum Optics Devices of Shanxi University (No. KF202006).

**Notes**

The authors declare no competing financial interest.